\documentclass[aps,pra,preprint,superscriptaddress,showpacs]{revtex4-1}
\usepackage{graphicx}
\usepackage{amsmath}
\usepackage{bm}

\begin{document}

\title{The influence of chiral spherical particles on the radiation of
optically active molecules}

\author{D.V. Guzatov}
\email[]{guzatov@gmail.com} \affiliation{Yanka Kupala Grodno State
University, Ozheshko str. 22, 230023 Grodno, Belarus}
\author{V.V. Klimov}
\email[]{vklim@sci.lebedev.ru} \affiliation{P.N. Lebedev Physical
Institute, Russian Academy of Sciences, Leninsky Prospect 53, 119991
Moscow, Russia}

\date{\today}

\begin{abstract}
In the framework of the perturbation theory of the nonrelativistic
quantum electrodynamics, a theory of spontaneous emission of a
chiral molecule located near a chiral (bi-isotropic) spherical
particle is developed. It is shown that the structure of photons in
the presence of chiral spherical particles differs significantly
from the structure of TE or TM photons. Exact analytical expressions
for the spontaneous emission radiative decay rate of a chiral
molecule with arbitrary electric and magnetic dipole moments of
transition located near a chiral spherical particle with arbitrary
parameters are obtained and analyzed in details. Simple asymptotes
for the case of a nanoparticle are obtained. Substantial influence
of even small chirality on a dielectric or ``left-handed'' sphere is
found. It is shown that by using chiral spherical particles it is
possible to control effectively the radiation of enantiomers of
optically active molecules.
\end{abstract}


\maketitle

\section{\label{sect1}Introduction}

Chirality is a geometric property of a three-dimensional body not to
coincide with its reflection in a mirror in any shifts and turns.
Such a property, for example, belongs to a human hand or a spring.
The term ``chirality'' was proposed by Lord Kelvin in 1873 to
explain some special properties of molecules \cite{ref1}. The
chirality plays an important role in biology and pharmacy, because
many complex organic compounds (amino acids, proteins, and sugars)
have chiral properties. For this reason, a body can react
differently with different enantiomers of the same substance. For
example, the same drug, depending on what type of molecules it
contains, may have a different taste and smell, or effect
differently. In physics, there is a considerable interest to chiral
media \cite{ref2,ref3,ref4,ref5}, in which there is a difference
between the distributions of left- and right-handed circularly
polarized electromagnetic waves. The simplest example of such an
optically active medium is an aqueous solution of sugar. At the
present time, the study of the chiral properties is again under
special attention because of the possibility to create metamaterials
based on chiral objects
\cite{ref6,ref7,ref8,ref9,ref10,ref11,ref12}.

The spontaneous emission of atoms and molecules in the vicinity of
material bodies is studied in a large number of works. The influence
of dielectric microspheres on the spontaneous emission of atoms is
considered in \cite{ref13,ref14,ref15,ref16}, spontaneous emission
of an atom located near a microsphere made of material with negative
refraction was considered in \cite{ref17} for first time. Optical
properties of chiral spherical particles have also been studied
quite extensively. At the present time, the scattering of plane
electromagnetic waves on homogeneous \cite{ref3,ref4,ref5} and
inhomogeneous \cite{ref18} chiral microspheres, including
multilayered chiral microspheres \cite{ref19,ref20}, is already
considered. The scattering of a Hermitian laser beam on a chiral
microsphere is studied in \cite{ref21}, the scattering of a plane
electromagnetic wave by a chiral sphere located in a chiral medium
is considered in \cite{ref22}. The radiation pressure force acting
on a chiral spherical particle placed in circularly polarized wave
is examined in \cite{ref23}. Finally, the radiation of a point
dipole source located inside a chiral spherical particle was
regarded in \cite{ref24}. At the same time, as far as we know, there
is no investigation devoted to the influence of chiral spherical
particles on the radiation of chiral (optically active) molecules.
However, such processes become more and more important because they
always appear when different biomolecules and drug molecules are
investigated and modified by optical methods. So, in this paper we
will present the detailed investigation of this problem

\begin{figure}[there]
\includegraphics[width=10.0cm]{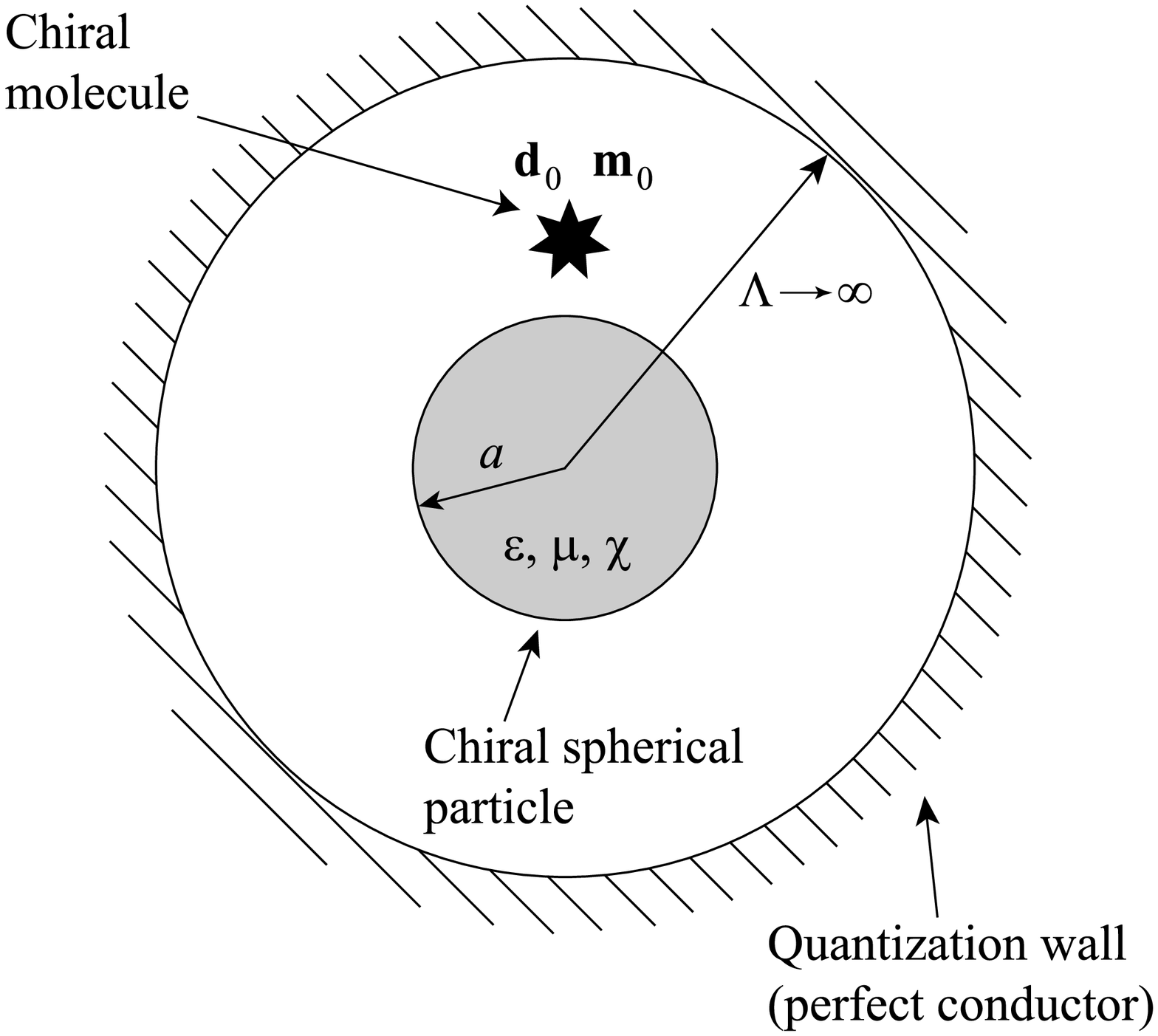}
\caption{\label{fig1} The geometry of the problem of spontaneous
emission of chiral molecules located near a chiral spherical
particle.}
\end{figure}

The goal of the present work is to develop an approach that allows
describing quantitatively the influence of a chiral spherical
particle made of an arbitrary metamaterial on the spontaneous
emission of a chiral molecule. All results will be obtained in the
framework of nonrelativistic quantum electrodynamics, in the
assumption that the radiative linewidth of the molecule is much
smaller than the frequency of radiation. The last assumption allows
us to use the perturbation theory. The rest part of the paper is
organized as follows. In the Section~\ref{sect2}, the canonical
quantization of the electromagnetic field in the presence of a
chiral spherical particle with arbitrary permittivity and
permeability is performed. In the Section~\ref{sect3}, we obtain
general expressions for the spontaneous emission radiative decay
rate of a chiral molecule located near a chiral sphere. In the
Section~\ref{sect4}, the graphical illustration of the results
obtained and their discussion are presented. The geometry of the
problem is shown in Fig.~\ref{fig1}.

\section{\label{sect2}Quantization of the electromagnetic field in the presence of a
chiral spherical particle}

To solve the problem of quantization of the electromagnetic field in
the presence of a chiral spherical particle, let us consider a
spherical cavity of an infinite radius $\Lambda \to \infty$ with a
perfectly conducting wall and with the a chiral particle located in
its center. For simplicity, we assume that the nanoparticle is
surrounded by vacuum (see Fig.~\ref{fig1}). The usual procedure of
quantization of the electromagnetic field in spherical geometry
\cite{ref25,ref26,ref27} is inapplicable in the case of chiral
particles, and below we will develop a new approach that is valid in
this case.

In the description of classical fields in the presence of chiral
spherical particles, we will use the method proposed in
\cite{ref4,ref28}. At the same time, to describe the chiral medium,
we use the constitutive equations in the Fedorov's form
\cite{ref29}:

\begin{eqnarray}
\label{eq1}
 \mathbf{D} &=& \varepsilon \left( {\mathbf{E} + \eta \textrm{rot} \mathbf{E}} \right), \nonumber \\
 \mathbf{B} &=& \mu \left( {\mathbf{H} + \eta \textrm{rot} \mathbf{H}} \right),
 \end{eqnarray}

\noindent where $\mathbf{D}$, $\mathbf{E}$ and $\mathbf{B}$,
$\mathbf{H}$ are the inductions and the strengths of electric and
magnetic fields, correspondingly, $\varepsilon$, $\mu$ are the
dielectric permittivity and the magnetic permeability of the chiral
medium, and $\eta$ is the dimensional parameter of chirality.
Substituting equations (\ref{eq1}) into Maxwell's equations, one can
obtain

\begin{eqnarray}
\label{eq2}
 &&\textrm{rot} \mathbf{E} - i\mu \chi \textrm{rot} \mathbf{H} = ik_{0} \mu \mathbf{H}, \nonumber \\
 &&\textrm{rot} \mathbf{H} + i\varepsilon \chi \textrm{rot} \mathbf{E} = - ik_{0} \varepsilon
\mathbf{E},
 \end{eqnarray}

\noindent where $k_{0} = \omega / c$ is the wave number in vacuum,
and $\chi = k_{0} \eta$ is the dimensionless parameter of chirality.
In Eq.~(\ref{eq2}) and everywhere further, the time dependence $\exp
\left( { - i\omega t} \right)$ is assumed. If we rewrite the system
(\ref{eq2}) in the matrix form:

\begin{equation}
\label{eq3}
\left( {{\begin{array}{*{20}c}
 {\textrm{rot} \mathbf{E}} \hfill \\
 {\textrm{rot} \mathbf{H}} \hfill \\
\end{array}}}  \right) = K\left( {{\begin{array}{*{20}c}
 {\mathbf{E}} \hfill \\
 {\mathbf{H}} \hfill \\
\end{array}}}  \right),\quad K = {\frac{{k_{0}}} {{1 - \chi ^{2}\varepsilon
\mu}} }\left( {{\begin{array}{*{20}c}
 {\chi \varepsilon \mu}  \hfill & {i\mu}  \hfill \\
 { - i\varepsilon}  \hfill & {\chi \varepsilon \mu}  \hfill \\
\end{array}}}  \right),
\end{equation}

\noindent then, with the help of the Bohren's transformation for the
electric and magnetic fields \cite{ref28}:

\begin{equation}
\label{eq4}
\left( {{\begin{array}{*{20}c}
 {\mathbf{E}} \hfill \\
 {\mathbf{H}} \hfill \\
\end{array}}}  \right) = A\left( {{\begin{array}{*{20}c}
 {\mathbf{Q}_{L}}  \hfill \\
 {\mathbf{Q}_{R}}  \hfill \\
\end{array}}}  \right),\quad A = \left( {{\begin{array}{*{20}c}
 {1} \hfill & { - i\mu / \sqrt {\mu \varepsilon}}   \hfill \\
 { - i\sqrt {\varepsilon \mu}  / \mu}  \hfill & {1} \hfill \\
\end{array}}}  \right),
\end{equation}

\noindent the matrix $K$ can be diagonalized:

\begin{equation}
\label{eq5}
A^{ - 1}KA = \left( {{\begin{array}{*{20}c}
 {k_{L}}  \hfill & {0} \hfill \\
 {0} \hfill & { - k_{R}}  \hfill \\
\end{array}}}  \right),
\end{equation}

\noindent where

\begin{equation}
\label{eq6}
k_{L} = {\frac{{k_{0} \sqrt {\varepsilon \mu}} } {{1 - \chi \sqrt
{\varepsilon \mu}} } },\quad k_{R} = {\frac{{k_{0} \sqrt {\varepsilon \mu}
}}{{1 + \chi \sqrt {\varepsilon \mu}} } },
\end{equation}

\noindent are the wave numbers of the left- ($L$) and
right-polarized ($R$) waves. The components of the transformed field
(\ref{eq4}) satisfy the equations \cite{ref4,ref28}:

\begin{eqnarray}
\label{eq7}
 &&\textrm{rot} \mathbf{Q}_{L} = + k_{L} \mathbf{Q}_{L} ,\quad \textrm{div} \mathbf{Q}_{L} = 0, \nonumber \\
 &&\textrm{rot} \mathbf{Q}_{R} = - k_{R} \mathbf{Q}_{R} ,\quad \textrm{div} \mathbf{Q}_{R} = 0.
 \end{eqnarray}

Thus, the fundamental solutions of the Maxwell's equations in a chiral
medium are the left-polarized and right-polarized waves with different wave
numbers.

In the spherical geometry under consideration, in the case of an
arbitrary electromagnetic field, the expressions for
$\mathbf{Q}_{L}$ and $\mathbf{Q}_{R}$ can be written down as follows
(we will omit a multiplier $\exp \left( { - i\omega t} \right)$ here
and below):

\begin{eqnarray}
\label{eq8}
 \mathbf{Q}_{L} &=& {\sum\limits_{n = 1}^{\infty}  {{\sum\limits_{m = - n}^{n}
{A_{mn}^{\left( {L} \right)} \left( {\bm{N \psi} _{mn}^{\left( {L}
\right)} + \bm{M \psi} _{mn}^{\left( {L} \right)}} \right)}
}}} , \nonumber \\
 \mathbf{Q}_{R} &=& {\sum\limits_{n = 1}^{\infty}  {{\sum\limits_{m = - n}^{n}
{A_{mn}^{\left( {R} \right)} \left( {\bm{N \psi} _{mn}^{\left( {R}
\right)} - \bm{M \psi} _{mn}^{\left( {R} \right)}} \right)} }}} ,
 \end{eqnarray}

\noindent where $A_{mn}^{\left( {L} \right)}$ and $A_{mn}^{\left(
{R} \right)}$ are the coefficients of expansions. These coefficients
can be found with the help of additional conditions. Explicit
expressions for spherical vector harmonics $\bm{N \psi}
_{mn}^{\left( {J} \right)}$ and $\bm{M \psi} _{mn}^{\left( {J}
\right)}$ (where $J=R$, $L$) are given in the Appendix.

From Eq.~(\ref{eq8}), it follows that $\mathbf{Q}_{L}$ and
$\mathbf{Q}_{R} $ are expressed in terms of fixed combinations of
vector spherical harmonics $\bm{N \psi} _{mn}^{\left( {L} \right)} +
\bm{M \psi }_{mn}^{\left( {L} \right)}$ and $\bm{N \psi
}_{mn}^{\left( {R} \right)} - \bm{M \psi} _{mn}^{\left( {R}
\right)}$, which always have a nonzero component along the radius
(see the Appendix) and therefore never can be reduced to the usual
TE and TM fields.

In accordance with the general rules of quantization of the electromagnetic
field, its expansion in the complete system of eigenfunctions of the
classical problem (standing spherical waves) can be represented as follows
(the indices are omitted):

\begin{eqnarray}
\label{eq9}
 \mathbf{\hat {E}}\left( {\mathbf{r}} \right) &=& i{\sum\limits_{s}
{{\frac{{a_{s} \mathbf{e}\left( {s,\mathbf{r}} \right) - a_{s}^{\dag}
\mathbf{e}^{ *} \left( {s,\mathbf{r}} \right)}}{{\sqrt {2}}} }}} , \nonumber \\
 \mathbf{\hat {H}}\left( {\mathbf{r}} \right) &=& {\sum\limits_{s}
{{\frac{{a_{s} \mathbf{h}\left( {s,\mathbf{r}} \right) +
a_{s}^{\dag} \mathbf{h}^{ *} \left( {s,\mathbf{r}} \right)}}{{\sqrt
{2}}} }}} ,
\end{eqnarray}

\noindent where $a_{s}$ and $a_{s}^{\dag}$ are the operators of
annihilation and creation of a photon in $s$-th mode respectively,
with usual rules of the commutation, $\mathbf{e}\left(
{s,\mathbf{r}} \right)$ and $\mathbf{h}\left( {s,\mathbf{r}}
\right)$ are the modes of the electric and the magnetic field, the
vector index $s = \left( {n,m,\nu}  \right)$ denotes a chosen set of
quantum numbers: the orbital ($n$), the azimuthal ($m$), and the
radial ($\nu$), the asterisk denotes the operation of the complex
conjugation. In accordance with Eqs.~(\ref{eq4}) and (\ref{eq8}),
the fields inside the chiral sphere can be written as

\begin{eqnarray}
\label{eq10}
\mathbf{e}^{in}\left( {s,\mathbf{r}}\right)  &=&
A_{mn}^{\left( {L}\right)
}\left( {\bm{N \psi}_{mn}^{\left( {L}\right) }+\bm{M \psi}_{mn}^{\left( {L} \right) }}\right) -i{\frac{{\mu }}{\sqrt{\varepsilon \mu }}} A_{mn}^{\left( {R}\right) }\left( {\bm{N \psi}_{mn}^{\left( {R}\right) }- \bm{M \psi}_{mn}^{\left( {R}\right) }}\right) ,  \nonumber \\
\mathbf{h}^{in}\left( {s,\mathbf{r}}\right)
&=&-{\frac{\sqrt{\varepsilon \mu }}{{\mu }}}A_{mn}^{\left(
{L}\right) }\left( {\bm{N \psi }_{mn}^{\left( {L}\right) }+\bm{M
\psi}_{mn}^{\left( {L}\right) }}\right) -iA_{mn}^{\left( {R}\right)
}\left( {\bm{N \psi}_{mn}^{\left( {R}\right) }-\bm{M \psi }
_{mn}^{\left( {R}\right) }}\right) .
\end{eqnarray}

Outside of the chiral particle, all waves have the same propagation velocity
and can be represented as

\begin{eqnarray}
\label{eq11}
 \mathbf{e}^{out}\left( {s,\mathbf{r}} \right) &=& C_{mn}^{\left( {1} \right)}
\bm{N \zeta} _{mn}^{\left( {1} \right)} + D_{mn}^{\left( {1}
\right)} \bm{M \zeta} _{mn}^{\left( {1} \right)} + C_{mn}^{\left(
{2} \right)} \bm{N \zeta} _{mn}^{\left( {2} \right)} +
D_{mn}^{\left( {2} \right)} \bm{M \zeta
}_{mn}^{\left( {2} \right)} , \nonumber \\
 \mathbf{h}^{out}\left( {s,\mathbf{r}} \right) &=& - \left( {C_{mn}^{\left(
{1} \right)} \bm{M \zeta} _{mn}^{\left( {1} \right)} +
D_{mn}^{\left( {1} \right)} \bm{N \zeta} _{mn}^{\left( {1} \right)}
+ C_{mn}^{\left( {2} \right)} \bm{M \zeta }_{mn}^{\left( {2}
\right)} + D_{mn}^{\left( {2} \right)} \bm{N \zeta} _{mn}^{\left(
{2} \right)}}  \right),
\end{eqnarray}

\noindent where $C_{mn}^{\left( {1} \right)}$, $C_{mn}^{\left( {2}
\right)} $ and $D_{mn}^{\left( {1} \right)}$, $D_{mn}^{\left( {2}
\right)} $ are some coefficients, and expressions for the spherical
vector harmonics $\bm{N \zeta} _{mn}^{\left( {j} \right)}$ and
$\bm{M \zeta} _{mn}^{\left( {j} \right)}$ (where $j = 1$, 2) are
given in the Appendix.

To find the coefficients in Eq.~(\ref{eq11}), one should use the
condition of continuity of tangential components of the electric and
magnetic fields on the surface of a chiral spherical particle having
the radius $a$. Since the molecule is situated outside the particle,
we do not need to know explicit values of the coefficients
$A_{mn}^{\left( {L} \right)}$ and $A_{mn}^{\left( {R} \right)}$ (see
Eq.~(\ref{eq8})). For the coefficients that describe fields outside
of the sphere, one can obtain the following relations from the
boundary conditions:

\begin{eqnarray}
\label{eq12}
 C_{mn}^{\left( {1} \right)} &=& \alpha _{n} C_{mn}^{\left( {2} \right)} +
\beta _{n} D_{mn}^{\left( {2} \right)} , \nonumber \\
 D_{mn}^{\left( {1} \right)} &=& \beta _{n} C_{mn}^{\left( {2} \right)} +
\gamma _{n} D_{mn}^{\left( {2} \right)} ,
 \end{eqnarray}

\noindent in which

\begin{eqnarray}
\label{eq13}
 \alpha _{n} = 1 - &&2\left( {{\frac{{A_{n} \left( {L} \right)V_{n} \left( {R}
\right) + A_{n} \left( {R} \right)V_{n} \left( {L} \right)}}{{W_{n} \left(
{L} \right)V_{n} \left( {R} \right) + W_{n} \left( {R} \right)V_{n} \left(
{L} \right)}}}} \right), \nonumber \\
 \beta _{n} = &&2\left( {{\frac{{W_{n} \left( {L} \right)A_{n} \left( {R}
\right) - W_{n} \left( {R} \right)A_{n} \left( {L} \right)}}{{W_{n} \left(
{L} \right)V_{n} \left( {R} \right) + W_{n} \left( {R} \right)V_{n} \left(
{L} \right)}}}} \right), \nonumber \\
 \gamma _{n} = 1 - &&2\left( {{\frac{{W_{n} \left( {L} \right)B_{n} \left( {R}
\right) + W_{n} \left( {R} \right)B_{n} \left( {L} \right)}}{{W_{n}
\left( {L} \right)V_{n} \left( {R} \right) + W_{n} \left( {R}
\right)V_{n} \left( {L} \right)}}}} \right),
 \end{eqnarray}

In Eq.~(\ref{eq13}), we have used the abbreviations ($J=L$, $R$):

\begin{eqnarray}
W_{n}\left( {J}\right) &=& P\psi _{n}\left( {k_{J}a}\right) \zeta _{n}^{\left( {1}\right) }{}^{\prime }\left( {k_{0}a}\right) -{\psi }_{n}^{\prime }\left( {k_{J}a}\right) \zeta _{n}^{\left( {1}\right) }\left( {k_{0}a}\right) , \nonumber \\
V_{n}\left( {J}\right) &=& \psi _{n}\left( {k_{J}a}\right) \zeta _{n}^{\left( {1}\right) }{}^{\prime }\left( {k_{0}a}\right) -P{\psi }_{n}^{\prime }\left( {k_{J}a}\right) \zeta _{n}^{\left( {1}\right) }\left( {k_{0}a}\right) , \nonumber \\
A_{n}\left( {J}\right) &=& P\psi _{n}\left( {k_{J}a}\right) {\psi
}_{n}^{\prime }\left( {k_{0}a}\right) -{\psi }_{n}^{\prime }\left(
{k_{J}a}\right) \psi
_{n}\left( {k_{0}a}\right) , \nonumber \\
B_{n}\left( {J}\right) &=& \psi _{n}\left( {k_{J}a}\right) {\psi
}_{n}^{\prime}\left( {k_{0}a}\right) -P{\psi }_{n}^{\prime }\left(
{k_{J}a}\right) \psi_{n}\left( {k_{0}a}\right) , \label{eq14}
\end{eqnarray}

\noindent where $P = \sqrt {\varepsilon \mu}  / \mu $, $\psi _{n}
\left( {x} \right)$ and $\zeta _{n}^{\left( {1} \right)} \left( {x}
\right)$ are the Riccati-Bessel functions (see the Appendix), and
the prime near the function means the derivative with respect to its
argument.

To study the interaction between a two-level molecule and the continuum of
electromagnetic modes modified by the presence of a chiral spherical
particle, it is necessary to know the density of final states. To find it,
one should use the condition of disappearance of tangential components of
electric field on the inner surface of the cavity. As a result we have

\begin{eqnarray}
C_{mn}^{\left( {1}\right) }\zeta _{n}^{\left( {1}\right) }{}^{\prime }\left( {k_{0}\Lambda }\right) +C_{mn}^{\left( {2}\right) }\zeta _{n}^{\left( {2} \right) }{}^{\prime }\left( {k_{0}\Lambda }\right) &=& 0, \nonumber \\
D_{mn}^{\left( {1}\right) }\zeta _{n}^{\left( {1}\right) }\left(
{k_{0}\Lambda }\right) +D_{mn}^{\left( {2}\right) }\zeta
_{n}^{\left( {2}\right) }\left( {k_{0}\Lambda }\right) &=& 0,
\label{eq15}
\end{eqnarray}

\noindent where $\zeta _{n}^{\left( {2} \right)} \left( {x} \right)$
is the Riccati-Bessel function (see the Appendix), and the prime
near the function means its derivative.

Substituting Eq.~(\ref{eq12}) into Eq.~(\ref{eq15}) and using
asymptotic expressions for the Hankel function at $\Lambda \to
\infty$ \cite{ref30}, we obtain

\begin{eqnarray}
\label{eq16}
 \left( {\alpha _{n} - \left( { - 1} \right)^{n + 1}\exp \left( { - 2ik_{0}
\Lambda}  \right)} \right)C_{mn}^{\left( {2} \right)} + \beta _{n}
D_{mn}^{\left( {2} \right)} &=& 0, \nonumber \\
 \beta _{n} C_{mn}^{\left( {2} \right)} + \left( {\gamma _{n} + \left( { -
1} \right)^{n + 1}\exp \left( { - 2ik_{0} \Lambda}  \right)}
\right)D_{mn}^{\left( {2} \right)} &=& 0.
 \end{eqnarray}

The system (\ref{eq16}) has nontrivial solutions only if its
determinant equals to zero. As a result, we have a quadratic
equation, which determines two different relationships between the
coefficients $C_{mn}^{\left( {2} \right)}$ and $D_{mn}^{\left( {2}
\right)}$

\begin{equation}
\label{eq17}
{\frac{{D_{mn}^{\left( {2} \right)}}} {{C_{mn}^{\left( {2} \right)}}} } = -
{\frac{{1}}{{2\beta _{n}}} }\left( {\alpha _{n} + \gamma _{n} \mp \sqrt
{\left( {\alpha _{n} + \gamma _{n}}  \right)^{2} - 4\beta _{n}^{2}}}
\right).
\end{equation}

The relations (\ref{eq17}) define two types of electromagnetic modes
(photons) occurring in the presence of a chiral spherical particle,
which are analogous to TM and TE modes in the case of a nonchiral
spherical particle. The upper sign ($-$) in Eq.~(\ref{eq17})
corresponds to modes which we will call $A$-type photons, while
another sign ($+$) gives us $B$-type photons. If the chirality
parameter tends to zero ($\chi \to 0)$, then, as it follows from
Eqs.~(\ref{eq13}) and (\ref{eq14}), the right side of
Eq.~(\ref{eq17}) tends to zero or to infinity, which leads to a
decoupling between TE and TM modes (vector spherical harmonics
$\bm{N \zeta} _{mn}^{\left( {2} \right)}$ and $\bm{M \zeta}
_{mn}^{\left( {2} \right)}$ in Eq.~(\ref{eq11})), and as a result
leads to the appearance of usual TM or TE polarized photons.

Thus, in the case of a chiral spherical particle we have two types
of polarization ($A$ and $B$), which are different from the TM and
TE polarizations.

Substituting Eq.~(\ref{eq17}) into Eq.~(\ref{eq16}), we find the
next asymptotic expressions ($\Lambda \to \infty$):

\begin{eqnarray}
\label{eq18}
 \left( {{\frac{{\omega _{s}}} {{c}}}} \right)_{A} &\approx& {\frac{{\pi
}}{{\Lambda}} }\left( {\nu + {\frac{{n + 1}}{{2}}} + {\frac{{i}}{{2\pi
}}} \textrm{Ln} \left( {\alpha _{n} + \beta _{n} O_{n}}  \right)} \right) + \ldots \nonumber \\
 \left( {{\frac{{\omega _{s}}} {{c}}}} \right)_{B} &\approx& {\frac{{\pi
}}{{\Lambda}} }\left( {\nu + {\frac{{n}}{{2}}} + {\frac{{i}}{{2\pi
}}} \textrm{Ln} \left( {\gamma _{n} + \beta _{n} O_{n}}  \right)}
\right) + \ldots
\end{eqnarray}

\noindent where $\omega _{s}$ is the mode eigenfrequency, and $c$ is
the speed of light in vacuum, and

\begin{equation}
\label{eq19}
O_{n} = - {\frac{{1}}{{2\beta _{n}}} }\left( {\alpha _{n} + \gamma _{n} -
\sqrt {\left( {\alpha _{n} + \gamma _{n}}  \right)^{2} - 4\beta _{n}^{2}}}
\right).
\end{equation}

From Eq.~(\ref{eq18}), it follows immediately that the density of
final states for both $A$-type and $B$-type photons has the same
value:

\begin{equation}
\label{eq20}
\rho \left( {\omega}  \right) = {\frac{{d\nu}} {{d\left( {\hbar \omega _{s}
} \right)}}} = {\frac{{\Lambda}} {{\pi \hbar c}}},
\end{equation}

\noindent and does not dependent on the presence of a spherical
particle of a final volume in the cavity, which is consistent with
the Courant's theorem \cite{ref31}.

For the normalization of modes, we will assume that there is only one photon
in the quantization volume. As a result, we obtain the condition:

\begin{equation}
\label{eq21}
{\frac{{1}}{{8\pi}} }{\int\limits_{V} {dV\left( {\mathbf{d}\left(
{s,\mathbf{r}} \right)\mathbf{e}^{ *} \left( {{s}',\mathbf{r}} \right) +
\mathbf{b}\left( {s,\mathbf{r}} \right)\mathbf{h}^{ *} \left(
{{s}',\mathbf{r}} \right)} \right)}}  = \delta _{s{s}'} \hbar \omega _{s} ,
\end{equation}

\noindent where $\delta _{s{s}'}$ is the Kronecker's delta, and the
inductions $\mathbf{d}$ and $\mathbf{b}$ of photonic modes can be
expressed via the operators of field strengths $\mathbf{e}$ and
$\mathbf{h}$ with the help of the constitutive equations
(\ref{eq1}). To calculate the integral in Eq.~(\ref{eq21}), we note
that the main contribution to it comes from the region $r \sim
\Lambda$ outside the spherical particle, where there is no chirality
and where it is possible to use asymptotic expressions for the
Hankel functions. As a result, we obtain ($n = 1$, 2, 3, $\ldots$;
$m = 0$, $\pm 1$, $\pm 2$, $\ldots$, $\pm n$):

\begin{equation}
{\left\vert {\left( {C_{mn}^{\left( {2}\right) }}\right)
_{A}}\right\vert }^{2}={\left\vert {\left( {D_{mn}^{\left(
{2}\right) }}\right) _{B}} \right\vert }^{2}={\frac{{\hbar \omega
_{s}^{3}}}{{2\Lambda c}^{2}{\left( { 1+{\left\vert
{O_{n}}\right\vert }^{2}}\right) }}}{\frac{{\left( {2n+1} \right)
}}{{n\left( {n+1}\right) }}}{\frac{{\left( {n-m}\right) !}}{{\left(
{n+m}\right) !}}}.  \label{eq22}
\end{equation}

Thus, expression (\ref{eq11}), (\ref{eq12}), and (\ref{eq22}), together with the expression for the
density of the final states (\ref{eq20}), fully describe the quantized
electromagnetic field in the presence of a chiral spherical particle, and
allow to describe the interaction of this field with arbitrary atoms and
molecules. In the next section, we will apply these expressions to find the
radiative decay rate of the spontaneous emission of a chiral molecule.

\section{\label{sect3}Radiative decay rate of spontaneous emission of a chiral molecule
located near a chiral spherical particle}

The radiative decay rate of the spontaneous emission (the radiative
linewidth $\gamma$) of a chiral molecule located near a chiral
spherical particle can be found by making use of the Fermi's
``golden rule'' \cite{ref32}:

\begin{equation}
\label{eq23}
\gamma = {\frac{{2\pi}} {{\hbar}} }{\sum\limits_{final} {{\left|
{{\left\langle {initial} \right|}H_{int} {\left| {final} \right\rangle}}
\right|}^{2}}} \rho \left( {\omega}  \right),
\end{equation}

\noindent where summation is performed over all possible final
states, which include both types of photons ($A$ and $B$, see
Eq.~(\ref{eq17})), $\rho \left( {\omega}  \right)$ is the density of
the final states (\ref{eq20}), and $H_{int}$ is the interaction
Hamiltonian of a chiral molecules with the electromagnetic field. In
this work, we assume that the Hamiltonian is Hermitian, i.e. there
are no losses in the material of a particle. If there are losses,
there is a possibility of nonradiative transition from the excited
state to the ground state. The expression for the nonradiative part
of the spontaneous decay rate can be obtained within the framework
of the classical approach \cite{ref33} or by using the nonstandard
quantization scheme \cite{ref34,ref35,ref36,ref37}. In further
calculations, for brevity we will consider only one channel of the
decay of the initial state of the molecule (transition $e \to g$),
i.e. the two-level molecule. To take into account the possibility of
transition of the molecule into several states, it is enough to sum
the partial linewidths.

The Hamiltonian of interaction of a chiral molecule with the
electromagnetic field in Eq.~(\ref{eq23}) can be written down in the
next form \cite{ref38}:

\begin{equation}
\label{eq24}
H_{int} = - \left( {\mathbf{\hat {d}}\mathbf{\hat {E}}\left( {\mathbf{r}_{0}
} \right)} \right) - \left( {\mathbf{\hat {m}}\mathbf{\hat {H}}\left(
{\mathbf{r}_{0}}  \right)} \right),
\end{equation}

\noindent where $\mathbf{r}_{0}$ is the radius vector of the
molecule position, $\mathbf{\hat {d}}$ and $\mathbf{\hat {m}}$ are
operators of the electric and the magnetic dipole moments of the
molecule, and field operators $\mathbf{\hat {E}}$ and $\mathbf{\hat
{H}}$ are defined in Eq.~(\ref{eq9}). We suppose that the initial
state corresponds to the vacuum state of the field and to the
molecule in the excited state, ${\left\langle {initial} \right|} =
{\left\langle {e} \right|}{\left\langle {vac} \right|}$, and that
the final field state correspond the single photon (of $A$-type or
$B$-type) and the molecule in the ground state ${\left| {final}
\right\rangle}  = {\left| {1_{n,m,\nu}}   \right\rangle} {\left| {g}
\right\rangle}$. Besides, for definiteness we will consider a spiral
model of a molecule in which electrons are constrained to move along
a helical path. Substituting Eq.~(\ref{eq24}) into Eq.~(\ref{eq23}),
we obtain the following expression for the total radiative decay
rate of spontaneous emission of the molecule:

\begin{equation}
\label{eq25}
\gamma _{eg} = \gamma _{eg}^{A} + \gamma _{eg}^{B} ,
\end{equation}

\noindent where, for example, in the case of photons of $A$-type

\begin{equation}
\label{eq26}
\gamma _{eg}^{A} = {\frac{{\pi}} {{\hbar}} }\rho \left( {\omega}
\right){\sum\limits_{n,m} {{\left| {\left( {\mathbf{d}_{0}
\mathbf{e}_{A}^{out} \left( {n,m,\nu ,\mathbf{r}_{0}}  \right)} \right) -
\left( {\mathbf{m}_{0} \mathbf{h}_{A}^{out} \left( {n,m,\nu ,\mathbf{r}_{0}
} \right)} \right)} \right|}^{2}}} ,
\end{equation}

\noindent where ${\left\langle {e{\left| {\mathbf{\hat {d}}}
\right|}g} \right\rangle } = \mathbf{d}_{0}$ and ${\left\langle
{e{\left| {\mathbf{\hat {m}}} \right|}g} \right\rangle}  = -
i\mathbf{m}_{0}$ are dipole moments of the considered transition of
the molecule on the frequency $\omega \approx \omega _{s} $ (see
Eq.~(\ref{eq18})). The chosen phase difference between the electric
and magnetic dipoles is due to the fact that the magnetic moment
operator is purely imaginary. This definition also agrees with the
adopted spiral model of a molecule. For more complicated chiral
molecules, there can be a different phase relation between electric
and magnetic dipole matrix elements.

Let us, for clarity, call the molecules with parallel
$\mathbf{d}_{0}$ and $\mathbf{m}_{0}$ the ``right'' molecules, while
the molecules with antiparallel $\mathbf{d}_{0}$ and
$\mathbf{m}_{0}$ will be referred to as the ``left'' molecules.
Explicit expressions for the component of Eq.~(\ref{eq25}) for the
mode of photons of $B$-type one can obtain from Eq.~(\ref{eq26}) by
replacing the index $A \to B$.

Let us assume for definiteness that the molecule is located on the
z-axis of a Cartesian coordinate system at the point $r_{0} > a$. In
this case, only components with $m = 0$, $\pm 1$ are nonzero, and
the explicit expressions for Eq.~(\ref{eq25}) will have the
following form:

\begin{equation}
\label{eq27}
\gamma _{eg}^{A} = \gamma _{eg}^{A, - 1} + \gamma _{eg}^{A,1} + \gamma
_{eg}^{A,0} ,
\end{equation}

\noindent where

\begin{eqnarray}
\gamma _{eg}^{A,-1}=&&{\frac{{k_{0}}}{{2\hbar
r_{0}^{2}}}}{\sum\limits_{n=1}^{\infty
}{{\frac{{2n+1}}{{1+{\left\vert {O_{n}}\right\vert }^{2}}}}{
\left\vert {\left( {d_{0x}-id_{0y}}\right) \left( {{\psi
}_{n}^{\prime }\left( {k_{0}r_{0}}\right) +T_{n}^{A}\zeta
_{n}^{\left( {1}\right)
}{}^{\prime }\left( {k_{0}r_{0}}\right) }\right) }\right. }}} \nonumber \\
&&-O_{n}\left( {d_{0y}+id_{0x}}\right) \left( {\psi _{n}\left( {k_{0}r_{0}} \right) +L_{n}^{A}\zeta _{n}^{\left( {1}\right) }\left( {k_{0}r_{0}}\right) } \right) \nonumber  \\
&&+O_{n}\left( {m_{0x}-im_{0y}}\right) \left( {{\psi }_{n}^{\prime
}\left( {k_{0}r_{0}}\right) +L_{n}^{A}\zeta _{n}^{\left( {1}\right)
}{}^{\prime}\left( {k_{0}r_{0}}\right) }\right) \nonumber  \\
&&{\left. {-\left( {m_{0y}+im_{0x}}\right) \left( {\psi _{n}\left(
{k_{0}r_{0}} \right) +T_{n}^{A}\zeta _{n}^{\left( {1}\right) }\left(
{k_{0}r_{0}}\right) } \right) }\right\vert }^{2}, \label{eq28}
\end{eqnarray}

\begin{eqnarray}
\gamma _{eg}^{A,1}=&&{\frac{{k_{0}}}{{2\hbar
r_{0}^{2}}}}{\sum\limits_{n=1}^{\infty
}{{\frac{{2n+1}}{{1+{\left\vert {O_{n}}\right\vert
}^{2}}}}{\left\vert {O_{n}\left( {d_{0y}-id_{0x}}\right) \left(
{\psi _{n}\left( {
k_{0}r_{0}}\right) +L_{n}^{A}\zeta _{n}^{\left( {1}\right) }\left( {k_{0}r_{0}}\right) }\right) }\right. }}} \nonumber \\
&&-\left( {d_{0x}+id_{0y}}\right) \left( {{\psi }_{n}^{\prime
}\left( {k_{0}r_{0}}\right) +T_{n}^{A}\zeta _{n}^{\left(
{1}\right)}{}^{\prime}\left( {k_{0}r_{0}}\right) }\right) \nonumber  \\
&&+\left( {m_{0y}-im_{0x}}\right) \left( {\psi _{n}\left(
{k_{0}r_{0}}\right) +T_{n}^{A}\zeta _{n}^{\left( {1}\right) }\left(
{k_{0}r_{0}}\right) }\right) \nonumber \\
&&{\left. {-O_{n}\left( {m_{0x}+im_{0y}}\right) \left( {{\psi
}_{n}^{\prime }\left( {k_{0}r_{0}}\right) +L_{n}^{A}\zeta
_{n}^{\left( {1}\right) }{}^{\prime }\left( {k_{0}r_{0}}\right)
}\right) }\right\vert }^{2}, \label{eq29}
\end{eqnarray}

\noindent and

\begin{eqnarray}
\label{eq30}
 \gamma _{eg}^{A,0} = &&{\frac{{2}}{{\hbar k_{0} r_{0}^{4}}} }{\sum\limits_{n
= 1}^{\infty}  {{\frac{{\left( {2n + 1} \right)n\left( {n + 1} \right)}}{{1
+ {\left| {O_{n}}  \right|}^{2}}}}{\left| {d_{0z} \left( {\psi _{n} \left(
{k_{0} r_{0}}  \right) + T_{n}^{A} \zeta _{n}^{\left( {1} \right)} \left(
{k_{0} r_{0}}  \right)} \right)} \right.}}} \nonumber  \\
 &&{\left. { + O_{n} m_{0z} \left( {\psi _{n} \left( {k_{0} r_{0}}  \right) +
L_{n}^{A} \zeta _{n}^{\left( {1} \right)} \left( {k_{0} r_{0}}
\right)} \right)} \right|}^{2}.
 \end{eqnarray}

In the case of photons mode of $B$-type, one can obtain

\begin{equation}
\label{eq31}
\gamma _{eg}^{B} = \gamma _{eg}^{B, - 1} + \gamma _{eg}^{B,1} + \gamma
_{eg}^{B,0} ,
\end{equation}

\noindent where

\begin{eqnarray}
\gamma _{eg}^{B,-1}=&&{\frac{{k_{0}}}{{2\hbar
r_{0}^{2}}}}{\sum\limits_{n=1}^{\infty
}{{\frac{{2n+1}}{{1+{\left\vert {O_{n}}\right\vert }^{2}}}}{
\left\vert {O_{n}\left( {d_{0x}-id_{0y}}\right) \left( {{\psi
}_{n}^{\prime }\left( {k_{0}r_{0}}\right) +T_{n}^{B}\zeta
_{n}^{\left( {1}\right)
}{}^{\prime }\left( {k_{0}r_{0}}\right) }\right) }\right. }}} \nonumber \\
&&-\left( {d_{0y}+id_{0x}}\right) \left( {\psi _{n}\left(
{k_{0}r_{0}}\right) +L_{n}^{B}\zeta _{n}^{\left( {1}\right) }\left(
{k_{0}r_{0}}\right) }\right) \nonumber \\
&&+\left( {m_{0x}-im_{0y}}\right) \left( {{\psi }_{n}^{\prime
}\left( {k_{0}r_{0}}\right) +L_{n}^{B}\zeta _{n}^{\left( {1}\right)
}{}^{\prime
}\left( {k_{0}r_{0}}\right) }\right) \nonumber  \\
&&{\left. {-O_{n}\left( {m_{0y}+im_{0x}}\right) \left( {\psi
_{n}\left( {k_{0}r_{0}}\right) +T_{n}^{B}\zeta _{n}^{\left(
{1}\right) }\left( {k_{0}r_{0}}\right) }\right) }\right\vert }^{2},
\label{eq32}
\end{eqnarray}

\begin{eqnarray}
\gamma _{eg}^{B,1}=&&{\frac{{k_{0}}}{{2\hbar
r_{0}^{2}}}}{\sum\limits_{n=1}^{ \infty
}{{\frac{{2n+1}}{{1+{\left\vert {O_{n}}\right\vert }^{2}}}}{
\left\vert {\left( {d_{0y}-id_{0x}}\right) \left( {\psi _{n}\left( {
k_{0}r_{0}}\right) +L_{n}^{B}\zeta _{n}^{\left( {1}\right) }\left( { k_{0}r_{0}}\right) }\right) }\right. }}} \nonumber \\
&&-O_{n}\left( {d_{0x}+id_{0y}}\right) \left( {{\psi }_{n}^{\prime
}\left( { k_{0}r_{0}}\right) +T_{n}^{B}\zeta _{n}^{\left( {1}\right)
}{}^{\prime
}\left( {k_{0}r_{0}}\right) }\right) \nonumber  \\
&&+O_{n}\left( {m_{0y}-im_{0x}}\right) \left( {\psi _{n}\left( {k_{0}r_{0}} \right) +T_{n}^{B}\zeta _{n}^{\left( {1}\right) }\left( {k_{0}r_{0}}\right) } \right) \nonumber  \\
&&{\left. {-\left( {m_{0x}+im_{0y}}\right) \left( {{\psi
}_{n}^{\prime }\left( {k_{0}r_{0}}\right) +L_{n}^{B}\zeta
_{n}^{\left( {1}\right) }{}^{\prime }\left( {k_{0}r_{0}}\right)
}\right) }\right\vert }^{2}, \label{eq33}
\end{eqnarray}

\noindent and

\begin{eqnarray}
\label{eq34}
 \gamma _{eg}^{B,0} = &&{\frac{{2}}{{\hbar k_{0} r_{0}^{4}}} }{\sum\limits_{n
= 1}^{\infty}  {{\frac{{\left( {2n + 1} \right)n\left( {n + 1} \right)}}{{1
+ {\left| {O_{n}}  \right|}^{2}}}}{\left| {O_{n} d_{0z} \left( {\psi _{n}
\left( {k_{0} r_{0}}  \right) + T_{n}^{B} \zeta _{n}^{\left( {1} \right)}
\left( {k_{0} r_{0}}  \right)} \right)} \right.}}} \nonumber  \\
 &&{\left. { + m_{0z} \left( {\psi _{n} \left( {k_{0} r_{0}}  \right) +
L_{n}^{B} \zeta _{n}^{\left( {1} \right)} \left( {k_{0} r_{0}}
\right)} \right)} \right|}^{2}.
 \end{eqnarray}

In Eqs.~(\ref{eq28})-(\ref{eq30}), (\ref{eq32})-(\ref{eq34}), we
have used the following notations:

\begin{eqnarray}
\label{eq35}
 T_{n}^{A} &=& {\frac{{1}}{{4}}}\left( {\alpha _{n} - \gamma _{n} + \sqrt
{\left( {\alpha _{n} + \gamma _{n}}  \right)^{2} - 4\beta _{n}^{2}}  - 2}
\right), \nonumber \\
 L_{n}^{A} &=& {\frac{{1}}{{4}}}\left( {\gamma _{n} - \alpha _{n} - \sqrt
{\left( {\alpha _{n} + \gamma _{n}}  \right)^{2} - 4\beta _{n}^{2}}  - 2}
\right), \nonumber \\
 T_{n}^{B} &=& {\frac{{1}}{{4}}}\left( {\alpha _{n} - \gamma _{n} - \sqrt
{\left( {\alpha _{n} + \gamma _{n}}  \right)^{2} - 4\beta _{n}^{2}}  - 2}
\right), \nonumber \\
 L_{n}^{B} &=& {\frac{{1}}{{4}}}\left( {\gamma _{n} - \alpha _{n} + \sqrt
{\left( {\alpha _{n} + \gamma _{n}}  \right)^{2} - 4\beta _{n}^{2}}
- 2} \right).
 \end{eqnarray}

In the case when the chirality parameter of a spherical particle is
equal to zero, Eqs.~(\ref{eq28})-(\ref{eq30}),
(\ref{eq32})-(\ref{eq34}) are simplified and have the following
form:

\begin{eqnarray}
\gamma _{eg}^{A,-1}=&&{\frac{{k_{0}}}{{2\hbar
r_{0}^{2}}}}{\sum\limits_{n=1}^{\infty }{\left( {2n+1}\right)
{\left\vert {\left( {d_{0x}-id_{0y}}\right) \left( {{\psi
}_{n}^{\prime }\left( {k_{0}r_{0}}\right) +T_{n}^{A}\zeta
_{n}^{\left( {1}\right) }{}^{\prime }\left( {k_{0}r_{0}}\right) }\right) } \right. }}} \nonumber \\
&&{\left. {-\left( {m_{0y}+im_{0x}}\right) \left( {\psi _{n}\left(
{k_{0}r_{0}} \right) +T_{n}^{A}\zeta _{n}^{\left( {1}\right) }\left(
{k_{0}r_{0}}\right) } \right) }\right\vert }^{2}, \label{eq36}
\end{eqnarray}

\begin{eqnarray}
\gamma _{eg}^{A,1}=&&{\frac{{k_{0}}}{{2\hbar
r_{0}^{2}}}}{\sum\limits_{n=1}^{ \infty }{\left( {2n+1}\right)
{\left\vert {\left( {d_{0x}+id_{0y}}\right) \left( {{\psi
}_{n}^{\prime }\left( {k_{0}r_{0}}\right) +T_{n}^{A}\zeta
_{n}^{\left( {1}\right) }{}^{\prime }\left( {k_{0}r_{0}}\right) }\right) } \right. }}} \nonumber \\
&&{\left. {-\left( {m_{0y}-im_{0x}}\right) \left( {\psi _{n}\left(
{k_{0}r_{0}} \right) +T_{n}^{A}\zeta _{n}^{\left( {1}\right) }\left(
{k_{0}r_{0}}\right) } \right) }\right\vert }^{2}, \label{eq37}
\end{eqnarray}

\begin{equation}
\label{eq38}
\gamma _{eg}^{A,0} = {\frac{{2{\left| {d_{0z}}  \right|}^{2}}}{{\hbar k_{0}
r_{0}^{4}}} }{\sum\limits_{n = 1}^{\infty}  {\left( {2n + 1} \right)n\left(
{n + 1} \right){\left| {\psi _{n} \left( {k_{0} r_{0}}  \right) + T_{n}^{A}
\zeta _{n}^{\left( {1} \right)} \left( {k_{0} r_{0}}  \right)} \right|}^{2}}
},
\end{equation}

\noindent and

\begin{eqnarray}
\gamma _{eg}^{B,-1}=&&{\frac{{k_{0}}}{{2\hbar
r_{0}^{2}}}}{\sum\limits_{n=1}^{ \infty }{\left( {2n+1}\right)
{\left\vert {\left( {d_{0y}+id_{0x}}\right)
\left( {\psi _{n}\left( {k_{0}r_{0}}\right) +L_{n}^{B}\zeta _{n}^{\left( {1} \right) }\left( {k_{0}r_{0}}\right) }\right) }\right. }}} \nonumber \\
&&{\left. {-\left( {m_{0x}-im_{0y}}\right) \left( {{\psi
}_{n}^{\prime }\left( {k_{0}r_{0}}\right) +L_{n}^{B}\zeta
_{n}^{\left( {1}\right) }{}^{\prime }\left( {k_{0}r_{0}}\right)
}\right) }\right\vert }^{2}, \label{eq39}
\end{eqnarray}

\begin{eqnarray}
\gamma _{eg}^{B,1}=&&{\frac{{k_{0}}}{{2\hbar
r_{0}^{2}}}}{\sum\limits_{n=1}^{ \infty }{\left( {2n+1}\right)
{\left\vert {\left( {d_{0y}-id_{0x}}\right) \left( {\psi _{n}\left( {k_{0}r_{0}}\right) +L_{n}^{B}\zeta _{n}^{\left( {1} \right) }\left( {k_{0}r_{0}}\right) }\right) }\right. }}} \nonumber \\
&&{\left. {-\left( {m_{0x}+im_{0y}}\right) \left( {{\psi
}_{n}^{\prime }\left( {k_{0}r_{0}}\right) +L_{n}^{B}\zeta
_{n}^{\left( {1}\right) }{}^{\prime }\left( {k_{0}r_{0}}\right)
}\right) }\right\vert }^{2}, \label{eq40}
\end{eqnarray}

\begin{equation}
\label{eq41}
\gamma _{eg}^{B,0} = {\frac{{2{\left| {m_{0z}}  \right|}^{2}}}{{\hbar k_{0}
r_{0}^{4}}} }{\sum\limits_{n = 1}^{\infty}  {\left( {2n + 1} \right)n\left(
{n + 1} \right){\left| {\psi _{n} \left( {k_{0} r_{0}}  \right) + L_{n}^{B}
\zeta _{n}^{\left( {1} \right)} \left( {k_{0} r_{0}}  \right)} \right|}^{2}}
},
\end{equation}

\noindent where

\begin{eqnarray}
T_{n}^{A}&=&-\left( {{\frac{{P\psi _{n}\left(
{k_{0}a\sqrt{\varepsilon \mu }}\right) {\psi }_{n}^{\prime }\left(
{k_{0}a}\right) -{\psi }_{n}^{\prime }\left(
{k_{0}a\sqrt{\varepsilon \mu }}\right) \psi _{n}\left( {k_{0}a}
\right) }}{{P\psi _{n}\left( {k_{0}a\sqrt{\varepsilon \mu }}\right)
\zeta _{n}^{\left( {1}\right) }{}^{\prime }\left( {k_{0}a}\right)
-{\psi } _{n}^{\prime }\left( {k_{0}a\sqrt{\varepsilon \mu }}\right)
\zeta
_{n}^{\left( {1}\right) }\left( {k_{0}a}\right) }}}}\right) , \nonumber \\
L_{n}^{B}&=&-\left( {{\frac{{\psi _{n}\left(
{k_{0}a\sqrt{\varepsilon \mu }} \right) {\psi }_{n}^{\prime }\left(
{k_{0}a}\right) -P{\psi }_{n}^{\prime }\left(
{k_{0}a\sqrt{\varepsilon \mu }}\right) \psi _{n}\left( {k_{0}a}
\right) }}{{\psi _{n}\left( {k_{0}a\sqrt{\varepsilon \mu }}\right)
\zeta _{n}^{\left( {1}\right) }{}^{\prime }\left( {k_{0}a}\right)
-P{\psi } _{n}^{\prime }\left( {k_{0}a\sqrt{\varepsilon \mu
}}\right) \zeta _{n}^{\left( {1}\right) }\left( {k_{0}a}\right)
}}}}\right) . \label{eq42}
\end{eqnarray}

Eqs.~(\ref{eq36})-(\ref{eq41}) correspond to components of the
radiative decay rate of spontaneous emission of a chiral molecule
located near the dielectric sphere. If we put $\mathbf{m}_{0} = 0$
or $\mathbf{d}_{0} = 0$ in these expressions, it will be possible to
find known relations for the radiative decay rate of spontaneous
emission of an atom located near a dielectric spherical particle,
and which has nonzero electric or magnetic dipole moment,
respectively \cite{ref39}.

Note, that despite of the fact, that this derivation was performed
in assumption of absence of losses in the material of the particle,
the expressions obtained will also exactly describe radiation losses
for nanoparticles made of real materials, i.e. for complex
$\varepsilon$ and $\mu$. It follows from the fact that classical and
quantum calculations give the same analytical expressions for the
radiative decay rate of spontaneous emission through the Green's
function of a classical problem \cite{ref40,ref41}. Since the
classical approach is valid in the case of both absorbing and
nonabsorbing bodies, this means that for a generalization of the
expressions obtained for the case of a real (absorbing) material of
a spherical particle, it is only necessary to substitute appropriate
complex values of dielectric permittivity and magnetic permeability
into obtained expressions.

To calculate the relative radiative decay rate of spontaneous
emission of chiral molecules located near a chiral spherical
particle, one should normalize the expression (\ref{eq25}) to the
value of the radiative decay rate of spontaneous emission of the
molecule in the absence of the particle. For this purpose, we use
Eqs.~(\ref{eq36})-(\ref{eq41}) where we put $T_{n}^{A} = L_{n}^{B} =
0$. By substituting the expressions obtained in such a way into
Eqs.~(\ref{eq27}), (\ref{eq31}) and (\ref{eq25}), and performing
summation, it is possible to find the radiative decay rate of
spontaneous emission of a chiral molecules in free space:

\begin{equation}
\label{eq43}
\gamma _{0,eg} = {\frac{{4k_{0}^{3}}} {{3\hbar}} }{\left| {\mathbf{d}_{0}}
\right|}^{2} + {\frac{{4k_{0}^{3}}} {{3\hbar}} }{\left| {\mathbf{m}_{0}}
\right|}^{2}.
\end{equation}

\noindent This expression, of course, coincides with the sum of the
known expressions for the linewidths of electric and magnetic
dipoles \cite{ref39}, because in the absence of chirality the
interference of radiation does not occur.

\section{\label{sect4}Analysis of the results obtained and illustrations}

Cumbersome formulas obtained in the previous section cannot fully reveal the
features of the spontaneous decay of a chiral molecule in the system under
consideration. In this section, we will explore the expressions found in the
typical special cases.

In a very important case of a chiral nanoparticle, i.e. a particle
having a size much smaller than the wavelength of the radiation, the
main contribution to Eqs.~(\ref{eq28})-(\ref{eq30}),
(\ref{eq32})-(\ref{eq34}) is from term with $n$ = 1. Expanding the
coefficients of Eq.~(\ref{eq13}) in a series of $k_{0} a \to 0$ and
keeping only the principal terms, we obtain

\begin{eqnarray}
\label{eq44}
 \alpha _{1} \approx 1 + &&{\frac{{4}}{{3}}}i\left( {k_{0} a}
\right)^{3}\left( {{\frac{{\left( {\varepsilon - 1} \right)\left( {\mu + 2}
\right) + 2\chi ^{2}\varepsilon \mu}} {{\left( {\varepsilon + 2}
\right)\left( {\mu + 2} \right) - 4\chi ^{2}\varepsilon \mu}} }} \right), \nonumber \\
 \beta _{1} \approx &&{\frac{{4}}{{3}}}i\left( {k_{0} a} \right)^{3}\left(
{{\frac{{3\chi \varepsilon \mu}} {{\left( {\varepsilon + 2} \right)\left(
{\mu + 2} \right) - 4\chi ^{2}\varepsilon \mu}} }} \right), \nonumber \\
 \gamma _{1} \approx 1 + &&{\frac{{4}}{{3}}}i\left( {k_{0} a}
\right)^{3}\left( {{\frac{{\left( {\mu - 1} \right)\left(
{\varepsilon + 2} \right) + 2\chi ^{2}\varepsilon \mu}} {{\left(
{\varepsilon + 2} \right)\left( {\mu + 2} \right) - 4\chi
^{2}\varepsilon \mu}} }} \right).
 \end{eqnarray}

The condition that the denominator of Eq.~(\ref{eq44}) will be a
zero

\begin{equation}
\label{eq45}
\left( {\varepsilon + 2} \right)\left( {\mu + 2} \right) - 4\chi
^{2}\varepsilon \mu = 0
\end{equation}

\noindent corresponds to the condition of a chiral-plasmon resonance
occurring in a spherical nanoparticle. The roots of Eq.~(\ref{eq45})
determine values of the constants $\varepsilon$ and $\mu$, that
correspond to a chiral plasmon. It should be noted, that the
resonant values of $\varepsilon$ and $\mu$ are not independent, and
their values differ from the corresponding values in the case of
plasmon resonance in a nonchiral particle, where (\ref{eq45}) is
split into two independent equations, $\varepsilon + 2 = 0$ and $\mu
+ 2 = 0$ \cite{ref42}.

To find an expression for the spontaneous emission radiative decay
rate in the case of a nanoparticle, one should expand $n = 1$ terms
in Eqs.~(\ref{eq28})-(\ref{eq30}), (\ref{eq32})-(\ref{eq34}) into a
series of $k_{0} a \to 0$ and keep only the principal terms.
However, it is a too long procedure, and here we will apply a
different approach used in \cite{ref43}. If the nanoparticle and the
molecule are placed close enough, then their total radiation will be
determined by their total electric and magnetic dipole moments:

\begin{equation}
\label{eq46}
\gamma _{eg} = {\frac{{4k_{0}^{3}}} {{3\hbar}} }{\left| {\mathbf{d}_{0} +
\delta \mathbf{d}} \right|}^{2} + {\frac{{4k_{0}^{3}}} {{3\hbar}} }{\left| {
- i\mathbf{m}_{0} + \delta \mathbf{m}} \right|}^{2},
\end{equation}

\noindent where $\delta \mathbf{d}$ and $\delta \mathbf{m}$ are the
electric and the magnetic dipole moments induced in the nanoparticle
while $\mathbf{d}_{0}$ and $ - i\mathbf{m}_{0}$ are the oscillating
dipole moments of the molecule. Explicitly, these values can be
written as follows \cite{ref43}:

\begin{eqnarray}
\label{eq47}
 \delta \mathbf{d} &=& \alpha _{EE} \mathbf{E}_{0} \left( {\mathbf{r}_{0}}
\right) + \alpha _{EH} \mathbf{H}_{0} \left( {\mathbf{r}_{0}}  \right), \nonumber \\
 \delta \mathbf{m} &=& \alpha _{HE} \mathbf{E}_{0} \left( {\mathbf{r}_{0}}
\right) + \alpha _{HH} \mathbf{H}_{0} \left( {\mathbf{r}_{0}}
\right),
 \end{eqnarray}

\noindent where the polarizability of a chiral spherical
nanoparticle in a homogeneous field is

\begin{eqnarray}
\label{eq48}
 \alpha _{EE} &=& a^{3}{\frac{{\left( {\varepsilon - 1} \right)\left( {\mu +
2} \right) + 2\chi ^{2}\varepsilon \mu}} {{\left( {\varepsilon + 2}
\right)\left( {\mu + 2} \right) - 4\chi ^{2}\varepsilon \mu}} }, \nonumber \\
 \alpha _{HH} &=& a^{3}{\frac{{\left( {\mu - 1} \right)\left( {\varepsilon +
2} \right) + 2\chi ^{2}\varepsilon \mu}} {{\left( {\varepsilon + 2}
\right)\left( {\mu + 2} \right) - 4\chi ^{2}\varepsilon \mu}} }, \nonumber \\
 \alpha _{EH} = - \alpha _{HE} &=& a^{3}{\frac{{3i\chi \varepsilon \mu
}}{{\left( {\varepsilon + 2} \right)\left( {\mu + 2} \right) - 4\chi
^{2}\varepsilon \mu}} }.
 \end{eqnarray}

The fields in Eq.~(\ref{eq47}) have the form:

\begin{equation}
\label{eq49} \mathbf{E}_{0} \left( {\mathbf{r}} \right) =
{\frac{{3\mathbf{n}\left( {\mathbf{n}\mathbf{d}_{0}}  \right) -
\mathbf{d}_{0}}} {{r^{3}}}},\quad \mathbf{H}_{0} \left( {\mathbf{r}}
\right) = - i{\frac{{3\mathbf{n}\left( {\mathbf{n}\mathbf{m}_{0}}
\right) - \mathbf{m}_{0}}} {{r^{3}}}},
\end{equation}

\noindent where $\mathbf{n}$ is the unit vector in the direction
from the center of the particle to the observation point.

In practice, the orientation of the molecule can be arbitrary with
respect to the nanoparticle surface, therefore to get an effective
radiative decay rate of spontaneous emission one should average
Eq.~(\ref{eq46}) on the orientations of molecules, or, equivalently,
on the unit vector $\mathbf{n}$. As a result we obtain

\begin{equation}
\gamma _{eg}^{eff}={\frac{{4k_{0}^{3}d}_{0}^{2}}{{3\hbar }}}{\left\{
{1+{\frac{{2}}{{r_{0}^{6}}}}{\left\vert {\alpha _{EE}-i\xi \alpha
_{EH}} \right\vert }^{2}+{\left\vert {\xi }\right\vert
}^{2}+{\frac{{2}}{{r_{0}^{6}}}}{\left\vert {\alpha _{HE}-i\xi \alpha
_{HH}}\right\vert }^{2}}\right\} }, \label{eq50}
\end{equation}

\noindent where $\xi$ is defined by the relation $\mathbf{m}_{0} =
\xi \mathbf{d}_{0}$.

As a rule, the magnetic dipole moment of the molecule is much
smaller than the electric dipole moment ${\left| {\mathbf{m}_{0}}
\right|} \ll {\left| {\mathbf{d}_{0}}  \right|}$. The chirality
parameter, even in hypothetical metamaterials, is also small ($\chi
\ll 1$). This fact determines that the second term in
Eq.~(\ref{eq50}), corresponding to the induced electric dipole
moment, is usually greater than the term corresponding to the
induced magnetic dipole moment. Thus, the effective interference
between the electric and magnetic fields is possible only if the
following two conditions take place:

\textit{1. In the system under consideration, a chiral-plasmon resonance must be present, i.e. the condition (\ref{eq45}) must be satisfied. Under this condition, the contribution of a magnetic radiation increases.}

\textit{2. The electric dipole moment induced in the nanoparticle
should be zero, i.e. the next condition must be satisfied (see
Eq.~(\ref{eq50}))}

\begin{equation}
\label{eq51}
\alpha _{EE} - i\xi \alpha _{EH} = 0.
\end{equation}

The solution of the system of Eqs.~(\ref{eq45}) and (\ref{eq51})
determines the values of dielectric permittivity and magnetic
permeability of the nanoparticle, which correspond to minimal values
of the radiative decay rate of a chiral molecule. It means that the
interference between the electric dipole and magnetic dipole
radiation becomes maximal and destructive if

\begin{eqnarray}
\label{eq52}
 \tilde {\mu}  &\to& - {\frac{{2d_{0}}} {{d_{0} + 2\chi m_{0}}} }, \nonumber \\
 \tilde {\varepsilon}  &\to& - {\frac{{2m_{0}}} {{m_{0} + 2\chi d_{0}}} }.
 \end{eqnarray}

From Eq.~(\ref{eq52}), it follows that by changing the sign in
$m_{0} $, i.e. when the chirality of the molecule is changed, the
``resonant'' magnetic permeability $\tilde {\mu}$ varies only
slightly ($\chi \ll 1$, $m_{0} \ll d_{0}$) and approximately equals
to $ - 2$. On the other hand, the ``resonant'' dielectric
permittivity $\tilde {\varepsilon}$ may have different signs for
molecules with different chirality. This means that both
nanoparticles with simultaneously negative $\varepsilon $ and $\mu$
(double negative - DNG-metamaterials), and nanoparticles with
negative $\mu$ and positive $\varepsilon$ ($\mu$ negative -
MNG-metamaterials or magnetic plasma) are suitable for the effective
control of radiation of chiral molecules. From the viewpoint of
practical implementation of such nanoparticles, the most suitable
are nanoparticles with a negative $\mu$ and positive $\varepsilon$,
because they can be implemented by using a well developed technology
of Split Rings Resonators \cite{ref44,ref45}. On the other, the
chiral DNG-metamaterials can be also synthesized \cite{ref12}.

The above expressions give a good description of properties of arbitrary
chiral molecules radiation of near nanoparticles, i.e., when the retardation
can be ignored. In the case when retardation effects are significant, it is
necessary to use the full expression for the decay rates given in the
previous section.

In any case, the process of spontaneous decay of a chiral molecule located
near a chiral spherical particle is very complex and below for clarity we
will give graphical illustrations of some possible regimes of interaction of
a chiral molecule with nanoparticles of different composition but neglecting
losses in them.

\begin{figure}[there]
\includegraphics[width=16.0cm]{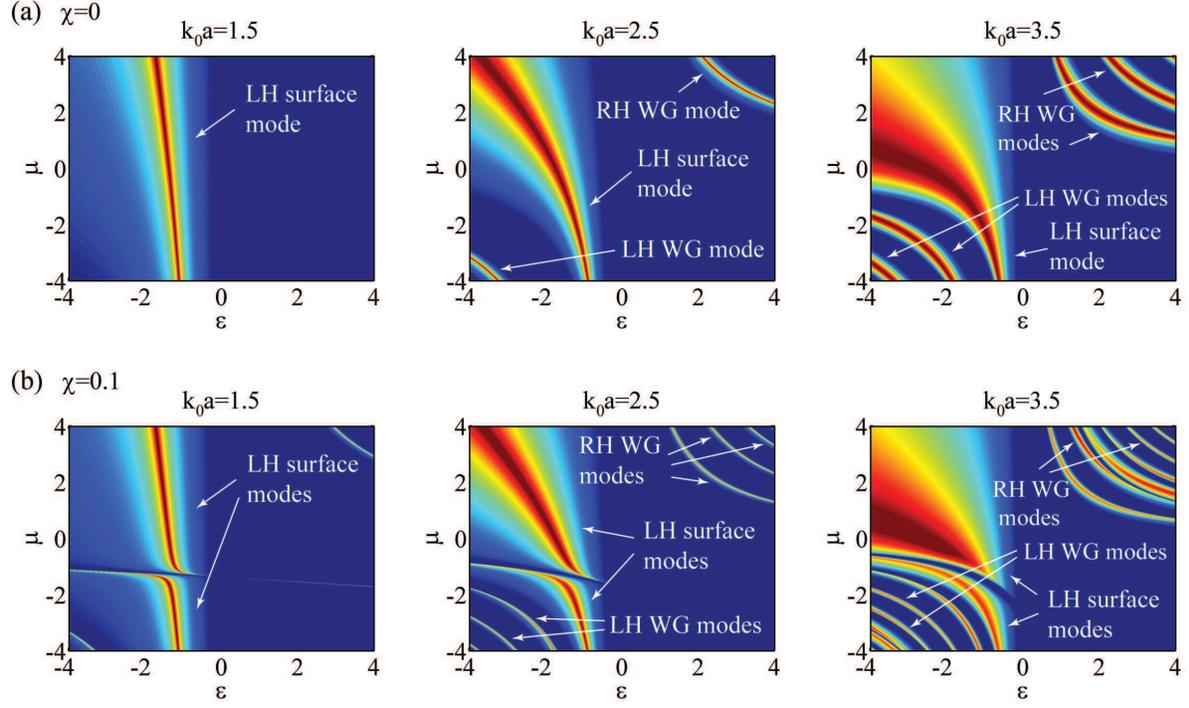}
\caption{\label{fig2} (Color online) The dependence of the absolute
values of the coefficient $T_{4}^{A}$ on the permittivity
($\varepsilon )$ and the permeability ($\mu )$ of chiral spherical
particle for different values of $k_{0} a$. (a) The chirality
parameter $\chi = 0$. (b) The chirality parameter $\chi = 0.1$. The
particle is placed in vacuum.}
\end{figure}

The Figure~\ref{fig2} shows the distribution of absolute values of
the coefficient $T_{4}^{A}$ (see Eq.~(\ref{eq35})) depending on the
dielectric permittivity and the magnetic permeability of the
material of nonchiral (Fig.~\ref{fig2}(a)) and chiral
(Fig.~\ref{fig2}(b)) spherical particles for different values of
$k_{0} a$. Large values of this coefficient correspond to a
resonance. This figure clearly shows that there is a significant
difference between these cases, as reflected in the fact that
nonzero chirality of the particle leads both to an increase in the
number of observed modes (whispering gallery (WG) modes in
dielectric ``right-handed'' (RH) material and in ``left-handed''
(LH) material, LH surfaces modes \cite{ref17}), and to the change of
their structure as compared with a nonchiral particle. This fact is
due to the simultaneous presence of both left- and right-polarized
waves in a chiral particle, which in turn leads to excitation of
resonances analogous to TM and TE resonances of nonchiral sphere,
simultaneously. The Figure~\ref{fig2} is very important because it
allows one to qualitative understand the structure of the radiative
decay rates of spontaneous emission of a chiral molecule placed near
the different nanoparticles.

\begin{figure}[there]
\includegraphics[width=9.0cm]{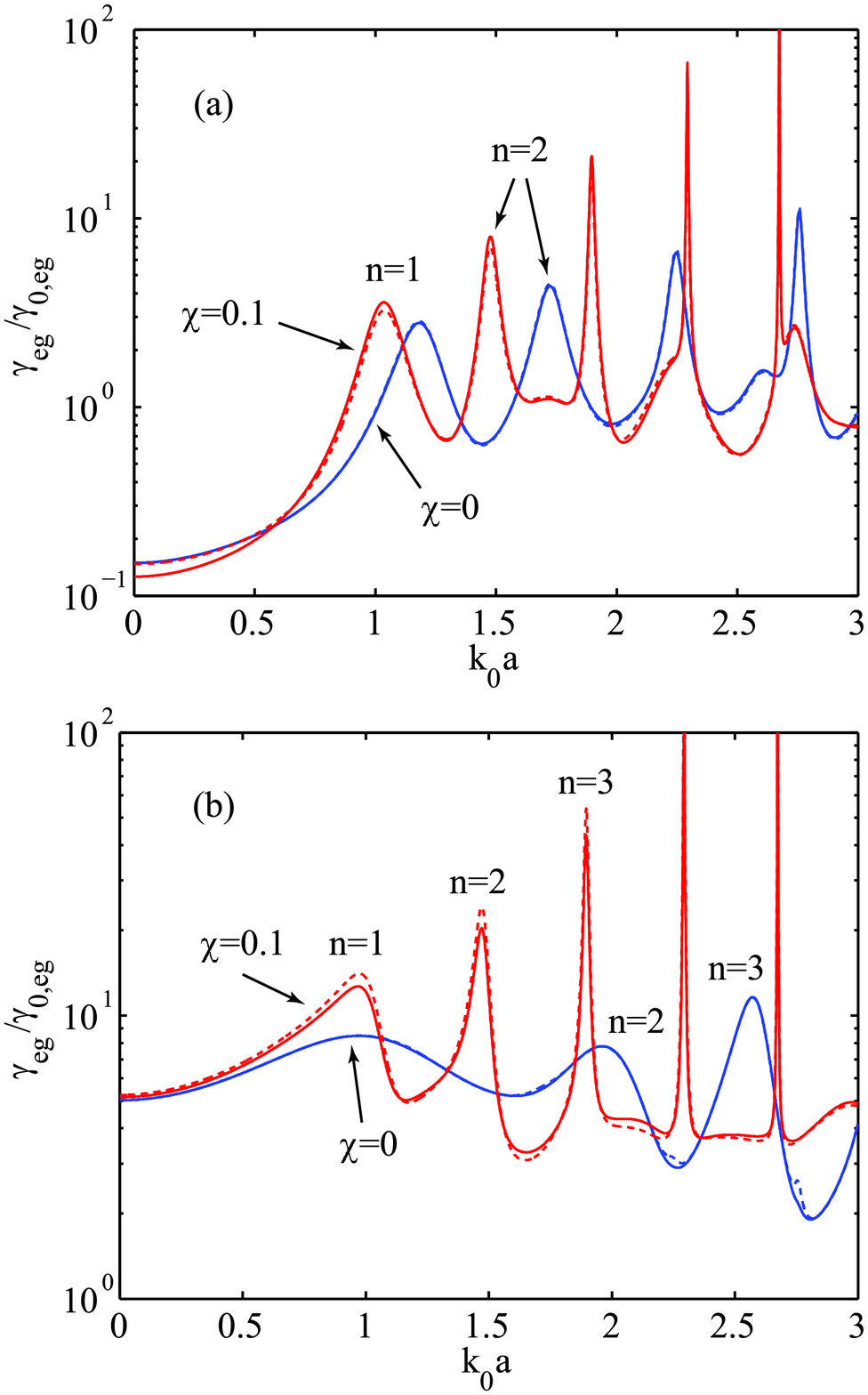}
\caption{\label{fig3} (Color online) Radiative decay rate of the
spontaneous emission of a chiral molecule located in close vicinity
to the surface of a chiral dielectric spherical particle ($r_{0} \to
a$) with $\varepsilon = 6$ and $\mu = 1$ as a function of its size,
$k_{0} a$. (a) The transition electric dipole moment of the molecule
is oriented along the $x$-axis ($\mathbf{d}_{0} = d_{0}
\mathbf{e}_{x}$), i.e. it is tangential to the surface of the
particle. (b) The transition electric dipole moment of the molecule
is oriented along the $z$-axis ($\mathbf{d}_{0} = d_{0}
\mathbf{e}_{z}$), i.e. it is normal to the surface of the particle.
The solid line corresponds to the transition magnetic dipole moment
of a molecule oriented along the x-axis, and the dashed line
corresponds to the transition magnetic dipole moment of a molecule
oriented along the $z$-axis. The value of the transition magnetic
dipole moment $\left\{ m_{0x},m_{0z}\right\} =0.1d_{0}$. The
particle is placed in vacuum.}
\end{figure}

It is very interesting that even for a dielectric sphere with a
small admixture of chirality, strong and unexpected effects occur.
In Fig.~\ref{fig3}, the radiative decay rate of spontaneous emission
of a chiral molecule placed near a chiral spherical particle with
positive values of permittivity and permeability is shown as a
function of the size parameter. As it is seen in this figure, the
presence of chirality in a spherical particle leads to a shift of
the maxima of the spontaneous emission radiative decay rate in
comparison with the case of a nonchiral particle. Indeed, as it is
known, a condition that allows to estimate the position of the
resonance is given by $k_{0} a \sqrt {\varepsilon \mu}  \approx n +
1 / 2$, where $n$ is the orbital quantum number. In the case of
chiral spherical particles, the left- and the right-polarized waves
exist simultaneously, and $k_{L} a > k_{0} a\sqrt {\varepsilon \mu}$
if $\chi > 0$. Hence the above condition should be changed to $k_{L}
a \approx n + 1 / 2$. As a result, this leads to smaller resonant
values of $k_{0} a$ than for a nonchiral particle (see
Fig.~\ref{fig2}). Even more interesting feature of ordinary
dielectric particles with a small admixture of chirality is a
substantial increase in the quality factor of whispering gallery
modes. This figure clearly shows that the corresponding linewidth
can be increased by the factor 9 or even more.

Note also that in Fig.~\ref{fig3}, only a slight difference between
the radiative decay rates of spontaneous emission corresponding to
the different orientations of the transition magnetic dipole moment
of a chiral molecule at a fixed electric dipole moment is observed.
However, one should expect a growth of this difference in the case
of a gradual increasing of absolute values of the magnetic dipole
momentum.

\begin{figure}[there]
\includegraphics[width=9.0cm]{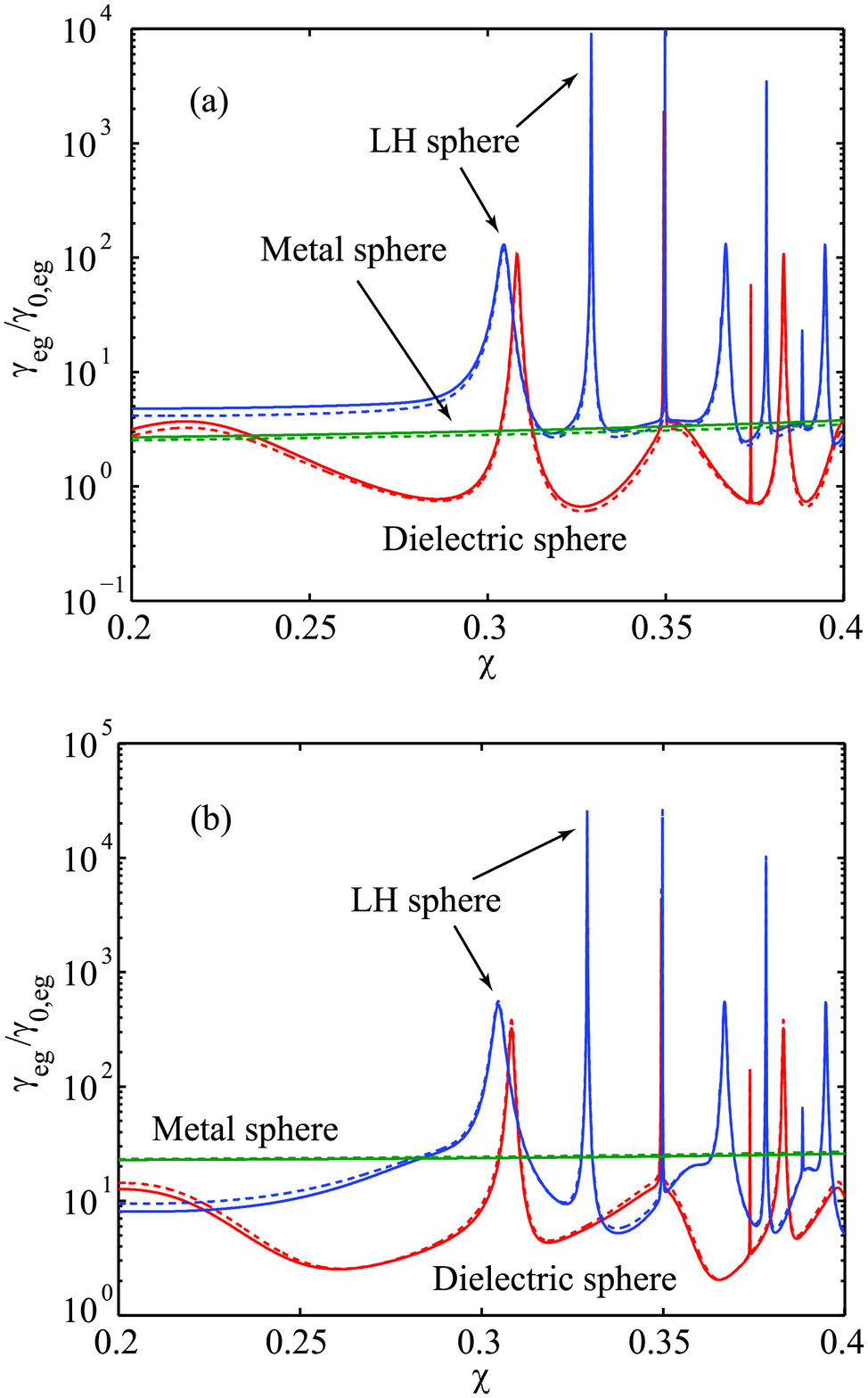}
\caption{\label{fig4} (Color online) The radiative decay rate of the
spontaneous emission of a chiral molecule located in close vicinity
to the surface of a chiral spherical particle ($r_{0} \to a$) with
the size $k_{0} a = 1$, as a function of the chirality parameter
$\chi $. A dielectric sphere has $\varepsilon = 4$ and $\mu = 1$, a
metal sphere has $\varepsilon = - 4$ and $\mu = 1$, an LH sphere has
$\varepsilon = - 4$ and $\mu = - 1.11$. (a) The transition electric
dipole moment of the molecule is oriented along the $x$-axis
($\mathbf{d}_{0} = d_{0} \mathbf{e}_{x}$), i.e. it is tangential to
the surface of the particle. (b) The transition electric dipole
moment of the molecule is oriented along the $z$-axis
($\mathbf{d}_{0} = d_{0} \mathbf{e}_{z}$), i.e. it is normal to the
surface of the particle. The solid line corresponds to the
transition magnetic dipole moment of the molecule oriented along the
x-axis, and the dashed line corresponds to the transition magnetic
dipole moment of the molecule oriented along the z-axis. The value
of the transition magnetic dipole moment $\left\{
m_{0x},m_{0z}\right\} =0.1d_{0}$. The particle is placed in vacuum.}
\end{figure}

Figure~\ref{fig4} shows the radiative decay rate of spontaneous
emission of a chiral molecule located near chiral spherical
nanoparticles ($k_{0} a = 1$) as a function of the chirality $\chi =
k_{0} \eta$. As it is clearly seen in this figure, changing the
chirality of the sphere has the greatest impact on the rate of
spontaneous decay of the molecules near dielectric and left-handed
spherical particles, in which high-Q modes can be excited. When the
chirality parameter approaches to the critical value $\chi _{crit} =
1 / \sqrt {\varepsilon \mu}$, then the number of oscillations in the
radiative decay rate increases rapidly, and $k_{L} \to \infty$ (see
Eq.~(\ref{eq6})). In the case of metallic particles, the dependence
of the radiative decay rate of spontaneous emission on the parameter
$\chi $, which is shown in Fig.~\ref{fig4}, is weak, due to
imaginary values of $\sqrt {\varepsilon \mu}$, and due to the
absence of propagating waves.

Chiral molecules play an especially important role in biology and
pharmacy. Therefore, it is extremely important to arrange the
separation of ``right'' and ``left'' enantiomers of molecules in the
racemic mixtures. This can be done in various ways, for example by
using radiation pressure forces in the electromagnetic field of
left- or right-handed circularly polarized electromagnetic waves
\cite{ref23}, or by using spiral optical beams \cite{ref46}. In
addition, such selection can be efficiently performed with the help
of chiral nanoparticles \cite{ref43}.

\begin{figure}[there]
\includegraphics[width=9.0cm]{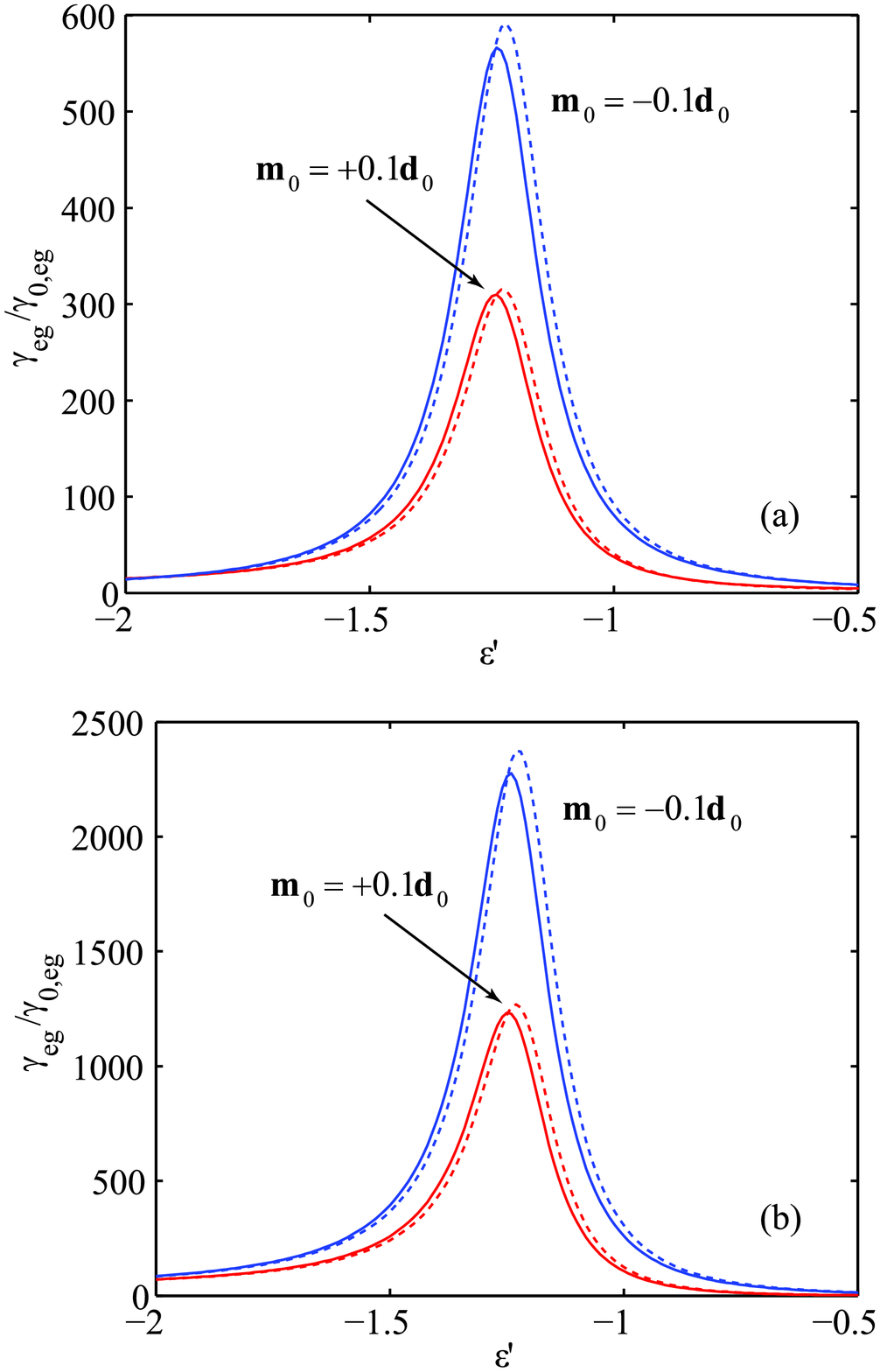}
\caption{\label{fig5} (Color online) The radiative decay rate of the
spontaneous emission of a chiral molecule located in close vicinity
to the surface of a chiral LH spherical nanoparticle ($r_{0} \to a$)
with $\varepsilon = {\varepsilon} ' + i0.1$, $\mu = - 1.6$, $\chi =
0.2$, and $k_{0} a = 0.1$. (a) The electric and magnetic dipole
moments of the transition of the molecule are oriented along the
$x$-axis (tangentially to the particle surface). (b) The electric
and magnetic dipole moments of the transition of the molecule are
oriented along the $z$-axis (normally to the particle surface). The
dashed line shows the asymptotic expression (\ref{eq46}). The
nanoparticle is placed in vacuum.}
\end{figure}

Figure~\ref{fig5} shows the radiative decay rate of spontaneous
emission of a chiral molecule located near the LH chiral spherical
nanoparticles as a function of the dielectric constant for a fixed
value of magnetic permeability. As it well seen in the figure, for
molecules that differ only in the orientation of the magnetic dipole
moment of the transition (the ``left'' molecule with $\mathbf{m}_{0}
= - 0.1\mathbf{d}_{0}$, and the ``right'' one with $\mathbf{m}_{0} =
+ 0.1\mathbf{d}_{0}$), the radiative decay rate takes significantly
different values, and thus, under such conditions it is possible to
control effectively the radiation properties of enantiomers. It is
important to note that both the results of calculations obtained in
the framework of the full QED theory (present paper), and in the
framework of a simplified quasistatic calculations presented at the
beginning of this section (see also \cite{ref42}) yield almost
identical results, confirming the correctness of both approaches.

\begin{figure}[there]
\includegraphics[width=9.0cm]{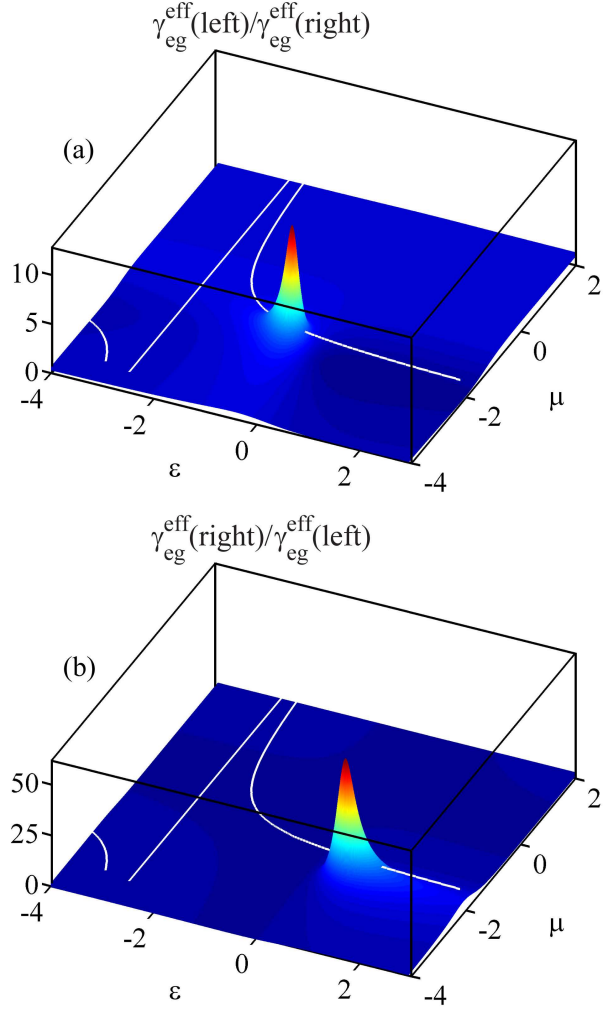}
\caption{\label{fig6} (Color online) (a) The ratio of the effective
radiative decay rate of spontaneous emission of ``left'' molecules
with $m_{0} = - 0.1d_{0}$ to the effective radiative decay rate of
spontaneous emission of ``right'' molecules with $m_{0} = +
0.1d_{0}$ and (b) vice versa as a function of the dielectric
permittivity ($\varepsilon$) and the magnetic permeability ($\mu$)
of the material from which chiral spherical nanoparticle is made.
The chirality parameter $\chi = 0.2$. The molecule is placed in
close vicinity to the surface of the spherical nanoparticle ($r_{0}
\to a$). The white line shows the position of chiral-plasmon
resonance in the nanoparticle (see Eq.~(\ref{eq45})). The
nanoparticle is placed in vacuum.}
\end{figure}

The Figure~\ref{fig6} shows the ratio of the effective radiative
decay rate of spontaneous emission of the ``left'' molecule and the
radiative decay rate of the ``right'' molecule and vice versa. From
this figure, it follows that if the condition (\ref{eq52}) is
satisfied, then the decay rates of ``left'' and ``right'' molecules
differ by the factor 15 or 60 or even more, depending on the
chirality of the molecule considered as a reference one. In other
words, nanoparticles with the parameter given by Eq.~(\ref{eq52})
will enhance the radiation of the ``right'' molecules and slow down
the radiation of the ``left'' molecules and vice versa. Let us
stress that ``left-handedness'' of a chiral sphere or its negative
$\mu$ are of crucial importance for such discrimination. Possible
applications of the effect of discrimination of the radiation one
can find in \cite{ref43}.

\section{\label{sect5}Conclusion}

Thus, in the present work the analytical expressions for the
radiative decay rate of spontaneous emission of a chiral molecule
located near an arbitrary spherical particle with chiral properties
were obtained and investigated within full QED theory. Using this
approach, the spontaneous emission decay rates of chiral molecule
placed near a spherical particle made of different materials
(dielectrics, metals, ``left-handed'' metamaterials etc.) were
investigated in details. It is shown that nonzero chirality of a
spherical particle leads to an increase in the number of excited
modes in comparison with nonchiral one due to coupling between left-
and right-polarized waves. Quasistatic expressions for the radiative
decay rate of spontaneous emission of chiral molecule located near a
chiral nanosphere were obtained and investigated. These expressions
are in good agreement with the exact QED results and allow one to
estimate the parameters of nanoparticles when the radiation of
``right'' or ``left'' molecules is suppressed. It was found that for
the suppression of the radiation of the ``left'' molecules the
sphere should be made of MNG metamaterials, while for the
suppression of radiation the ``right'' molecules the DNG materials
should be used.

The results obtained can be used to calculate the radiative decay rate of
spontaneous emission of chiral molecules in the vicinity of the chiral
spherical particles, for interpretation of experimental data on interaction
of chiral molecules and particles, and for detection and selection of chiral
molecules with the help of chiral particles.

\appendix*
\section{\label{appen}Spherical vector harmonics}

Spherical vector harmonics that describe the electromagnetic field
inside a chiral spherical particle have the form:

\begin{eqnarray}
\label{eq53}
 \bm{N \psi} _{mn}^{\left( {J} \right)} = &&n\left( {n + 1}
\right){\frac{{\psi _{n} \left( {k_{J} r} \right)}}{{\left( {k_{J}
r} \right)^{2}}}}P_{n}^{m} \left( {\cos \theta}  \right)e^{im\phi
}\mathbf{e}_{r} + {\frac{{{\psi} '_{n} \left( {k_{J} r}
\right)}}{{k_{J} r}}}{\frac{{\partial P_{n}^{m} \left( {\cos \theta}
\right)}}{{\partial
\theta}} }e^{im\phi} \mathbf{e}_{\theta} \nonumber  \\
 &&+ i{\frac{{{\psi} '_{n} \left( {k_{J} r} \right)}}{{k_{J}
r}}}{\frac{{mP_{n}^{m} \left( {\cos \theta}  \right)}}{{\sin \theta
}}}e^{im\phi} \mathbf{e}_{\phi}  ,
 \end{eqnarray}

\begin{eqnarray}
\label{eq54}
 \bm{M \psi}_{mn}^{\left( {J} \right)} = i{\frac{{\psi _{n}
\left( {k_{J} r} \right)}}{{k_{J} r}}}{\frac{{mP_{n}^{m} \left(
{\cos \theta } \right)}}{{\sin \theta}} }e^{im\phi}
\mathbf{e}_{\theta} - {\frac{{\psi _{n} \left( {k_{J} r}
\right)}}{{k_{J} r}}}{\frac{{\partial P_{n}^{m} \left( {\cos \theta}
\right)}}{{\partial \theta}} }e^{im\phi }\mathbf{e}_{\phi}  ,
 \end{eqnarray}

\noindent where $0 \le r < \infty$, $0 \le \theta < \pi$ and $0 \le
\phi < 2\pi$ are spherical coordinates, $\mathbf{e}_{r}$,
$\mathbf{e}_{\theta}$, $\mathbf{e}_{\phi} $ are unit vectors of the
spherical coordinate system, $\psi _{n} \left( {k_{J} r} \right) =
\left( {\pi k_{J} r / 2} \right)^{1 / 2}J_{n + 1 / 2} \left( {k_{J}
r} \right)$ is the Riccati-Bessel function \cite{ref30}, where $J_{n
+ 1 / 2} \left( {k_{J} r} \right)$ is the Bessel function
\cite{ref30}, the prime near the function means the derivative with
respect to its argument, $P_{n}^{m} \left( {\cos \theta}  \right)$
is the associated Legendre function \cite{ref30}, and the index
$J=L$, $R$.

Spherical vector harmonics that describe the electromagnetic field
outside a chiral spherical particle have the following form:

\begin{eqnarray}
\bm{N \zeta}_{mn}^{\left( {j}\right) }=&&n\left( {n+1}\right)
{\frac{{\zeta _{n}^{\left( {j}\right) }\left( {k_{0}r}\right)
}}{{\left( {k_{0}r}\right) ^{2}}}}P_{n}^{m}\left( {\cos \theta
}\right) e^{im\phi }\mathbf{e}_{r}+{\frac{{\zeta _{n}^{\left( {j}\right) }{}^{\prime }\left( {k_{0}r}\right) }}{{k_{0}r}}}{\frac{{\partial P_{n}^{m}\left( {\cos \theta }\right) }}{{\partial \theta }}}e^{im\phi }\mathbf{e}_{\theta } \nonumber \\
&&+i{\frac{{\zeta _{n}^{\left( {j}\right) }{}^{\prime }\left(
{k_{0}r}\right) }}{{k_{0}r}}}{\frac{{mP_{n}^{m}\left( {\cos \theta
}\right) }}{{\sin \theta }}}e^{im\phi }\mathbf{e}_{\phi },
\label{eq55}
\end{eqnarray}

\begin{eqnarray}
\label{eq56}
 \bm{M \zeta} _{mn}^{\left( {j} \right)} = i{\frac{{\zeta
_{n}^{\left( {j} \right)} \left( {k_{0} r} \right)}}{{k_{0}
r}}}{\frac{{mP_{n}^{m} \left( {\cos \theta}  \right)}}{{\sin \theta
}}}e^{im\phi} \mathbf{e}_{\theta} - {\frac{{\zeta _{n}^{\left( {j}
\right)} \left( {k_{0} r} \right)}}{{k_{0} r}}}{\frac{{\partial
P_{n}^{m} \left( {\cos \theta}  \right)}}{{\partial \theta}}
}e^{im\phi} \mathbf{e}_{\phi}  ,
 \end{eqnarray}

\noindent where $\zeta _{n}^{\left( {j} \right)} \left( {k_{0} r}
\right) = \left( {\pi k_{0} r / 2} \right)^{1 / 2}H_{n + 1 /
2}^{\left( {j} \right)} \left( {k_{0} r} \right)$ is the
Riccati-Bessel function \cite{ref30}, where $H_{n + 1 / 2}^{\left(
{j} \right)} \left( {k_{0} r} \right)$ is the Hankel function of the
first ($j = 1$) and second ($j = 2$) kind, respectively, and the
prime near the function means the derivative with respect to its
argument.

More information about the properties of spherical vector harmonics
can be found, for example, in \cite{ref47}.

\pagebreak

\end{document}